\documentclass[12pt]{report}
\usepackage{fullpage}
\usepackage{amssymb}
\usepackage{epsf}
\usepackage{multicol}
\usepackage{bbold}		
\newcommand{\Case}[2]{{\textstyle \frac{#1}{#2}}}
\begin{document}
\thispagestyle{empty}
$~$
\vskip 2.0cm 
\centerline{{\LARGE{\bf Lectures on Constrained Systems }}}
\vskip 2.0cm 
\centerline{{\Large{\bf Ghanashyam Date}}}
\vskip 1.0cm
\centerline{{{\large The Institute of Mathematical Sciences}}}
\vskip 0.5cm
\centerline{{\large C.I.T. Campus, Tharamani, Chennai, 600 113, India.}}
\vfill
\hrule
\vskip 0.5cm
\leftline{{email: shyam@imsc.res.in }}

\newpage
\centerline{{\LARGE{\bf Preface}} }
\vskip 0.75cm
These lecture notes have been prepared as a basic introduction to
the theory of constrained systems which is how basic forces of nature
appear in their Hamiltonian formulation. Only a basic familiarity of
Lagrangian and Hamiltonian formulation of mechanics is assumed. 

The First chapter makes some introductory remarks indicating the context
in which various types of constrained systems arise. It distinguishes
constrained systems for which the equations of motion are uniquely
specified from those for which the equations of motion are {\em not}
uniquely determined. The focus of the lectures is on the latter types of
constraints. The notations that will be used is introduced here.

In the second chapter, the general features are introduced in the familiar 
example of source-free electrodynamics.

In the third chapter, features seen in electrodynamics are abstracted in
the simpler context of systems with finitely many degrees of freedom.
How constraints arise in a Lagrangian and in a Hamiltonian formulation
is discussed.

In the fourth chapter, the Dirac's algorithm for discovering constraints
and their classification is discussed in a sufficiently general context.

In the fifth chapter, a special class of constrained systems for which the
Hamiltonian itself is a constraint is discussed. Such systems arise in
the context of general relativity and lead to a host of issues of
interpretation of `dynamics'.

A set of exercises is also included for practice in the last chapter. 

These set of eight lectures were given at the {\em Refresher Course for College
Teachers} held at IMSc during May-June, 2005. 

\vskip 0.75cm
June 4, 2005 \hfill Ghanashyam Date
\tableofcontents
\chapter{Introductory Remarks}

Let us begin by recalling some elementary understanding of what one
means by mechanics. There are two distinct parts: (a) kinematics and (b)
dynamics. Kinematics specifies the (possible) {\em states} of a given
system while dynamics specifies how a system evolves from one state to
another state. The central problem one wants to solve is to find the
state of system at some time $t$ given its state at an (earlier) time
$t_0$. While there are various types of `state spaces' possible, we will
be dealing with those systems for which the state space is a {\em
manifold} (finite or infinite dimensional) and an evolution which is of
the continuous variety.

There are three distinct tasks before us (i) devise a {\em framework}
or a calculational prescription which will take as input a state space,
some quantity reflecting/encoding a dynamics and give us a corresponding
rule to evolve any given state of the system. A familiar example is the
Lagrangian framework wherein the state space is described in terms of
positions and velocities, dynamics is encoded in terms of a Lagrangian
function and the variational principle leads to the Euler-Lagrange
equations of motion. (ii) Use physical, qualitative analysis of a given
physical system to associate a particular state space and a particular
Lagrangian (say) with the physical system and (iii) devise methods to
obtain generic evolutions in sufficiently explicit terms. 

The last task is where one will make `predictions' and is the most
relevant in applications (eg engineering). Here the issues such as
sensitivity to initial conditions etc play an important role. This task
can be addressed only after the first two are specified. This is {\em
not} the aspect we will discuss in these lectures. Some of it will be
discussed in the NLDL part of the course.

To be definite, let us consider the Lagrangian framework. Thus we will
have some {\em configuration space}, $Q$, which is some $n$ dimensional
manifold with coordinates $q^i, i = 1, 2, \cdots n$. The dynamics is
encoded in a function $L(q^i, \dot{q}^i, t)$ and the equations of motion
are the Euler-Lagrange equations of motion,
\begin{equation} \label{ELEqn}
\delta S ~=~ \delta \int dt L(q, \dot(q)) ~=~ 0 ~~\Rightarrow~~
\frac{d}{d t} \frac{\partial L}{\partial \dot{q}^i} - \frac{\partial
L}{\partial q^i} ~=~ 0, ~~~~ \forall ~ i.
\end{equation}

Questions: How general can $Q$ be? How general can $L$ be? Can we {\em
always} obtain equations of motion via the variational principle?

Generally, one takes $Q$ to be an n-dimensional manifold. However, this
{\em need not} be $\mathbb{R}^N$, although usually it is. For instance
one may begin with $n$ particles moving in 3 dimensions, so that $Q$ is
$3 n$ dimensional, but there could be restrictions such as distance
between any two particles is always fixed (rigid body) or that particles
must have velocities tangential to the surface of a ball. In the latter
case, the relevant $Q$ would be the 2-sphere which is a compact manifold
without boundary. In both cases the {\em dimension} of the relevant
configuration space is smaller that what we began with. We could also
imagine confining the particles to a 3 dimensional box in which case the
relevant configuration space is a bounded portion of $\mathbb{R}^{3n}$.
In short, the possible motion of the system may be {\em constrained}. A
majority of applied mechanics problems have to deal with such
constrained systems and there is a massive recent treatise on such
systems \cite{Treatise}. Typically, such constrained systems are
described by specifying relations $f_{\alpha}(t, \vec{r}, \vec{v}) = 0$.
These relations are required to be {\em functionally independent},
mutually consistent, and valid for {\em all possible forces}. Some
terminology that you might have come across classifies the constraints
as: (1) Relations independent of velocities are termed {\em
positional/holonomic/finite constraints}; (2) Velocity dependent ones
are called {\em velocity constraints} which are further divided into
{\em scleronomic} (time independent), {\em rheonomic} (time dependent),
{\em holonomic} (integrable), {\em non-holonomic} (non-integrable).
Constrained systems of this variety, serve to restrict the configuration
space and also possible motions, but the {\em Euler-Lagrange equations
are always solvable for accelerations, so that dynamics is uniquely
specified.} We will {\em not} be dealing with this type of constrained
systems.

The constrained systems that we will deal with arise in situations where
the well motivated choices of the configuration space and the
Lagrangian, do {\em not} specify the equations of motion uniquely. In
the Lagrangian framework, these correspond to the so called {\em
singular} Lagrangians. Analysis of such systems is better carried in the
Hamiltonian framework wherein one arrives at an understanding of {\em
gauge theories} and we will be essentially focusing on such {\em
constrained Hamiltonian systems}.

From the point of view of usual applications of classical mechanics,
such systems would appear quite exotic and possibly `irrelevant'. However 
{\em all} the four fundamental interactions that we know of, when cast in a
Lagrangian or Hamiltonian framework, {\em precisely} correspond to the
kind of constrained systems we will discuss. An understanding of these
basic forces at the classical structural level is crucial also for
constructing/understanding the corresponding quantum theories. In the
context of general relativity, the constrained nature of the theory
throws up challenging conceptual and interpretational issues. With these
reasons as primary motivations, we will discuss constrained systems.

We will also use certain notations which are introduced below.
\begin{enumerate}

\item I will freely use terms such as `manifold', `tensors' etc. Here is
a very rapid, heuristic introduction to these terms.

The idea of a {\em manifold} (actually a differentiable manifold), is
invented to be able to do {\em differential and integral calculus} on
arbitrarily complicated spaces such as (say) the surface of an
arbitrarily shaped balloon. The basic definition of differentiation
involves taking differences of values of a function at two nearby points
eg., $f(x + h) - f(x)$, dividing it by $h$ and taking the limit $h \to
0$. This is fine when $x, h$ etc are numbers. But when we go to the
surface of a balloon, we do not know how to implement such definitions
eg how to take the `difference of two points' on the surface. The way
the idea of differentiation is captured is to {\em assign} numbers to
points and then use them in the usual manner. This assignment can be
thought of as pasting small pieces of a graph paper on the surface, and
reading off the numbers corresponding to the points just below the
numbers. This pasting gives a set of {\em local coordinates} in that
portion of the surface. Clearly there is huge freedom in the choice of
the pieces of graph paper as well as in the pasting. Thus, while local
coordinates can be introduced, the arbitrariness must be respected.
Changing these assignments is called {\em local coordinate
transformations}. Likewise, quantities such as tiny arrows stuck on the
surface can be described in terms of {\em components} of the arrows with
respect to the coordinate axes provided by the graph paper, When the
graph paper is changed, these components change, but {\em not} the
arrows themselves. Consequently, the two sets of components must be
related to the local coordinate transformations in a specific manner so
that both the sets refer to the same arrow. The arrows are an example of
a {\em vector field} which can be thought of as a collection of their
components, transforming in a specific manner. The components are
denoted as quantities with arbitrary number of {\em upper and lower
indices}. These must transform as,
\begin{equation}
(X')^{i_1, \cdots, i_m}_{j_1, \cdots, j_n} (y(x)) ~ = ~
\frac{\partial y^{i_1}}{\partial x^{r_1}}\  \cdots\
\frac{\partial y^{i_m}}{\partial x^{r_m}}\  
\frac{\partial x^{s_1}}{\partial y^{j_1}}\  \cdots\
\frac{\partial x^{s_m}}{\partial y^{j_m}}\  
(X)^{r_1, \cdots, r_m}_{s_1, \cdots, s_n} (x)
\end{equation}
Objects represented by such indexed quantities have a meaning {\em
independent} of the choice of local coordinates and are called {\em
Tensors of contravariant rank $m$ and covariant rank $n$}. Equations
involving tensors (correctly matched index distribution) are {\em
covariant} (or form invariant) with respect to local coordinate
transformations. This is all that we will need to know.

\item In the context of Lagrangian framework, the generalized
coordinates will be denoted by $q^i$. These are not necessarily
Cartesian and are to be thought of as arbitrary local coordinates on the
configuration space manifold. Consequently, we will occasionally also
comment on whether various equations/conditions are
`covariant/invariant' under coordinate transformations. For these
purposes, tensor notation will be used freely. Lagrangian will always be
taken to be a function of generalized coordinates and velocities and
independent of time.

\item In the context of Hamiltonian framework, the generalized
{\em canonical} coordinates will be denoted as $q^i, p_i$. It will be
convenient to subsume them as $2n$ coordinates $\omega^{\mu}$. The phase
space will be denoted as $\Gamma$ which is a $2n$ dimensional manifold.
On this there is a distinguished antisymmetric rank 2 tensor,
$\Omega_{\mu\nu}(\omega)$ which satisfies further properties namely,
\begin{equation}\label{Closed}
\partial_{\lambda}\Omega_{\mu\nu} + \mathrm{cyclic} = 0.
\end{equation}
The antisymmetric matrix $\Omega_{\mu\nu}$ is assumed to be {\em
invertible} and its inverse is denoted as $\Omega^{\mu\nu}$,
$\Omega^{\mu\lambda}\Omega_{\lambda\nu} = \delta^{\mu}_{\nu}$.

Such an antisymmetric tensor is called a {\em symplectic form}.
The coordinates in which the matrix takes the block off-diagonal form
\begin{eqnarray}
\Omega^{\mu\nu} ~=~ 
\left( \begin{array}{cc} \mathbb{0} & \mathbb{1} \\ -\mathbb{1} &
\mathbb{0} \end{array} \right)
& , & 
\Omega_{\mu\nu} ~=~ 
\left( \begin{array}{cc} \mathbb{0} & -\mathbb{1} \\ \mathbb{1} &
\mathbb{0} \end{array} \right)
\end{eqnarray}
are the {\em canonical coordinates} and these are guaranteed to exist by
Darboux theorem.

\item The usual Poisson bracket of two functions $f(\omega), g(\omega)$
get expressed as,
\begin{equation}
\{f(\omega), g(\omega)\} ~=~ \Omega^{\mu\nu}\frac{\partial f}{\partial
\omega^{\mu}} \frac{\partial g}{\partial \omega^{\nu}}
\end{equation}
The (\ref{Closed}) condition implies the Jacobi identity of Poisson
brackets.

For every function $f$ on the phase space, one can associate the
so-called {\em Hamiltonian} vector field, $v_f^{\mu} :=
\Omega^{\mu\nu}\partial_{\nu} f$. Such Hamiltonian vector fields
generate {\em infinitesimal canonical transformations} as,
\begin{equation}
\omega^{\mu} \to \omega'^{\mu} := \omega^{\mu} + \epsilon
v^{\mu}_f(\omega).
\end{equation}
It is easy to check that the symplectic form is invariant under these
coordinate transformations (and hence these are called canonical
transformations). In particular, the Hamiltonian function generates
time dependent changes in coordinates which is nothing but the dynamical
evolution. Hence, dynamical evolution can be viewed as a ``continuous
unfolding of canonical transformations". Taking $\epsilon := \delta t, f
= H$ implies the Hamilton's equations of motion, 
\begin{equation}
\frac{d \omega^{\mu}}{d t} = \Omega^{\mu\nu} \frac{\partial H}{\partial
\omega^{\nu}}
\end{equation}
Note that these are not just any system of first order, ordinary
differential equation, but have a specific form involving the
antisymmetric tensor $\Omega^{\mu\nu}$ because of which $H$ is constant
along solutions of equations of motion. Furthermore, one also has the
Liouville theorem regarding conservation of phase space volumes.

\end{enumerate}

We will begin our discussion by taking the example of source free
electrodynamics.

\chapter{Hamiltonian Formulation of Electrodynamics}

We will begin with the usual Maxwell equations, put them in the four
dimensional relativistic form, arrive at an action formulation from
which we will go to the Lagrangian and Hamiltonian form. To be able to
use the four dimensional relativistic tensor notation, we need to choose
a set of conventions.

{\underline{Maxwell Equations:}}
\begin{eqnarray}
\vec{\nabla}\cdot \vec{E} ~=~ \rho &,& 
\vec{\nabla}\times \vec{E} + \frac{\partial \vec{B}}{\partial t} ~=~ 0
\\
\vec{\nabla}\cdot \vec{B} ~=~ 0 
&,&
\vec{\nabla}\times \vec{B} - \frac{\partial \vec{E}}{\partial t} ~=~
\vec{j} \\
\frac{\partial \rho}{\partial t} & = & - \vec{\nabla} \cdot \vec{j}
\end{eqnarray}

Conventions:
\begin{itemize}

\item Write all quantities without derivatives and with arrows as
quantities with {\em upper} indices i.e. as {\em contravariant tensor},
eg, $\vec{a} \leftrightarrow (a^1, a^2, a^3) \leftrightarrow a^i$ and
vector derivatives as derivatives with {\em lower} indices ({\em
covariant tensor}), eg, $\vec{\nabla} \leftrightarrow (\partial_1,
\partial_2, \partial_3) \leftrightarrow \partial_i$.

Thus, electric field, magnetic field and vector potential are
contravariant tensors while the gradient operator is a covariant
quantity. The cross product explicitly gives a contravariant quantity.

\item Quantities with upper indices are related to those with lower
indices by, $a^i = - a_i, a^0 = a_0$ i.e. $a^{\mu} :=
\eta^{\mu\nu}a_{\nu}$ with $\eta^{\mu\nu} = \mathrm{diag}(1, -1, -1, -1)
= \eta_{\mu\nu}$.

\item Cross products are expressed as, $(\vec{\nabla} \times \vec{a})^i
:= \epsilon^{ij}_{~~k} \partial_j a^k, \mathrm{where,} ~\epsilon^{123} =
1 = - \epsilon_{123}$. It then follows that,
\begin{equation} 
\epsilon^{ij}_{~~k} \epsilon^k_{~mn} = - (\delta^i_m \delta^j_n -
\delta^i_n \delta^j_m) \ .  
\end{equation}
\end{itemize}

Maxwell equations now become,
\begin{eqnarray}
\partial_i E^i ~ = ~ \rho &,& \epsilon^{ij}_{~~k} \partial_j B^k -
\frac{\partial E^i}{\partial t} ~=~ j^i \label{Source}\\
\partial_i B^i ~ = ~ 0 &,& \epsilon^{ij}_{~~k} \partial_j E^k +
\frac{\partial B^i}{\partial t} ~=~ 0 \label{SourceFree}
\end{eqnarray}

Define, $j^0 := \rho, ~ F^{0i} := - E^i, ~ F^{ij} := -
\epsilon^{ij}_{~~k} B^k$. It is easy to check that 
\begin{eqnarray} (\ref{Source}) 
& \to & \partial_{\nu} F^{\nu\mu} = j^{\mu} ~, \label{SourceTwo}\\
(\ref{SourceFree}) & \to &  \partial^{\mu} F^{\nu\lambda} +
\mathrm{cyclic} ~ = ~ 0 ~ = ~ \partial_{\mu}F_{\nu\lambda} +
\mathrm{cyclic} \label{SourceFreeTwo}
\end{eqnarray}

It follows that $B^i = \frac{\mathrm{1}}{\mathrm{2}} \epsilon^i_{~jk}
F^{jk}$ and the identification of the electric field matches with the
usual definition with $A^{\mu} \leftrightarrow (A^0 := \Phi, \vec{A})$.

This allows us to think of Maxwell equations as tensor equations
involving 4 dimensional tensors. If one transforms the coordinates
$x^{\mu}$ as $x'^{\mu} := \Lambda^{\mu}_{\nu} x^{\nu}$ and transforms
the $F, j$ quantities by tensor rules (index-by-index action), then
evidently the Maxwell equations are {\em form invariant} or {\em
covariant} for {\em all} invertible matrices $\Lambda^{\mu}_{\nu}$.
Remember though that we have also used specific rules to relate the
upper and the lower indexed quantities. These rules must also be
respected by the primed observer (coordinates) i.e. $\eta_{\mu\nu}$ must
also be {\em invariant} under the coordinate transformations and this
restricts the $\Lambda^{\mu}_{\nu}$ to be the familiar Lorentz
transformation matrices. We thus also see the Lorentz covariance of
Maxwell equations. Incidentally, operating by $\partial_{\mu}$ on
(\ref{SourceFreeTwo}) and using (\ref{SourceTwo}) one can deduce the
wave equation, $\Box F_{\mu\nu} = 0$ and invariance of $\Box :=
\eta^{\mu\nu}\partial_{\mu}\partial_{\nu}$ is precisely the requirement
of Lorentz invariance. 

Note that the 4-tensor notation, is just a compact notation to write
Maxwell equations and the physical property of electrodynamics being
Lorentz covariant is encoded by the invariance of $\Box$ which in turn
requires invariance of $\eta$. The compact notation helps to keep track
of the Lorentz covariance property.

Now we would like to see if these equations can be obtained from an
action principle and we all know that the answer is of course yes. For
our purposes, it is sufficient to consider the source free case and we
will now take $j^{\mu} = 0$.

If $F_{\mu\nu}$ are treated as basic variables describing
electrodynamics, then we have 8 first order partial differential
equations for six quantities. Usual equations of Lagrangian framework
are second order (in $t$) equations for configuration space variables.
However one notices that (\ref{SourceFreeTwo}) can be {\em identically}
solved by putting $F_{\mu\nu} := \partial_{\mu} A_{\nu} - \partial_{\nu}
A_{\mu}$. This is always possible to do locally. Now the remaining 4
equations become 4 second order equations for the 4 quantities
$A_{\mu}$. At least now the number of equations equals the number of
unknowns. However, the definition of $F$ in terms of $A$ does {\em not}
determine $A$ uniquely (not even up to a constant); $A'_{\mu} = A_{\mu}
+ \partial_{\mu} \Lambda$ gives the same $F$. Therefore, although we
have correct number of equations, these equations do not suffice to
determine the candidate dynamical variables, $A$, uniquely. Modulo this
observation, let us go ahead any way by thinking of $A_{\mu}(x^{\alpha})$
as `configurations space variables'.

It is a very easy exercise to check that if we define an action, 
\begin{equation}
S[A(x)] := \int dt \int d^3x \left( - \frac{1}{4} F_{\mu\nu} F^{\mu\nu}
\right) := \int d^4x {\cal L}(A_{\mu}, \partial_{\alpha} A_{\beta})\ ,
\end{equation}
then its stationarity condition gives the Maxwell equations expressed in
terms of $A$. This of course is an example of a {\em field system} i.e.
a dynamical system with {\em infinitely many degrees of freedom},
$A_{\mu}(t, x^i)$ with $\mu, x^i$ serving as labels. To proceed further,
let us introduce some working rules and notations: derivatives of the
Lagrangian density etc will be denoted as $\frac{\delta {\cal
L}(x)}{\delta A_{\mu}(x')}$ with the rules,
\begin{equation} 
\frac{\delta A_{\mu}(x)}{\delta A_{\nu}(x')} = \delta^{\nu}_{\mu}
\delta^3 (x^i - x'^i) ~~,~~ \frac{\delta \partial_{\nu}
A_{\mu}(x)}{\delta A_{\lambda}(x')} = \delta^{\lambda}_{\mu}
\partial_{\nu} \delta^3 (x^i - x'^i) ~~.  
\end{equation}

Now let us try to get to a Hamiltonian formulation. Our basic
configuration space variables are $A_{\mu}(t, x^i)$ while the Lagrangian
is $L = \int d^3x (-1/4) F_{\mu\nu}F^{\mu\nu}$. The conjugate momenta
are given by,
\begin{equation}
\pi^{\mu}(x) := \int d^3x' \frac{\delta {\cal L}(x)}{\delta \partial_0
A_{\mu}(x')} ~~=~~ - F^{0\mu}(x) \ .
\end{equation}
Clearly, $\pi^0 = 0$ identically while $\pi^i = - F^{0i}$. Therefore,
$\pi_i = -F_{0i} = - \partial_0A_i + \partial_i A_0$. This allows us to
express the `velocities' $\partial_0 A_i = - \pi_i + \partial_i A_0$
which is needed in getting to the Hamiltonian form. However, {\em we
cannot do so for $\partial_0 A_0$}!

The canonical Hamiltonian is defined as, \begin{eqnarray} H_c & = & \int
d^3x \left[\pi^{\mu}\partial_0 A_{\mu} + \frac{1}{4} F_{\mu\nu}
F^{\mu\nu}\right] \\ & = & \int d^3 x\left[ \pi^i \left( \partial_0 A_i
- \partial_i A_0 + \partial_i A_0 \right) - \frac{1}{2}
  \left(F^{0i}\right)^2 + \frac{1}{4} F_{ij} F^{ij} + \pi^0 \partial_0
  A_0 \right] \nonumber \\ & = & \int d^3 x\left[ -\frac{1}{2} \pi^i
  \pi_i + \frac{1}{4} F_{ij} F^{ij} - A_0 (\partial_i \pi^i ) + \pi^0
  \partial_0 A_0 \right] \end{eqnarray}
In the last equation, a partial integration has been done. The first two
terms are the usual electromagnetic field energy density, $\vec{E}\cdot
\vec{E} + \vec{B}\cdot \vec{B}$.  The last term depends only on $\pi^0,
\partial_0 A_0$ and is {\em not} a function of phase space variables, it
contains a velocity. If we naively consider the Hamilton's equation of
motion for $A_0$, we get an identity, implying that the time dependence
of $A_0$ is undetermined. The equation of motion for $\pi^0$ however
leads to $\partial_0 \pi_0 ~=~ \partial_i \pi^i $.  Observe that {\em
if} we drop the last term by setting $\pi^0 = 0$ at some initial time,
and {\em require} this condition to hold for all times, {\em then} we
will need $\partial_i \pi^i = 0$ for all time which is just one of the
Maxwell equations (the Gauss Law).  This is also suggested by the
Legendre transformation involved in going from the Lagrangian to the
Hamiltonian -- we are required to use the definition of the momenta.
Hence, in the Hamiltonian we should not have the last term. Since $\pi_0
= 0$ from definition {\em at all times}, we must also obtain $\partial_0
\pi_0 = 0$ from the Hamilton's equation of motion. Thus it is self
consistent to interpret the dropping of the last term as {\em imposition
of a constraint, $\pi^0 = 0$} on the phase space of $(A^{\mu},
\pi_{\mu})$ and the Gauss law following as a consistency condition.
Since $A_0$ remains undetermined, we will replace $A_0$ in the third
term by an {\em arbitrary} function of time, $\lambda(t)$.  Since the
Gauss law is also independent of positions/velocities, we will interpret
it also as a constraint, $\chi := \partial_i \pi^i$. We will refer to
$\pi^0 = 0$ as a {\em primary constraint} and $\chi = 0$ as a {\em
secondary constraint}, because it is needed for the primary constraint
to hold for all times. 

Let us consider now the canonical transformations generated by the two
constraints. It follows,
\begin{eqnarray}
\delta_{\epsilon} A_{\mu}(x) & = &\left\{A_{\mu}(x), \int d^3 x'
\epsilon(x') \pi^0(x')\right\} ~~=~~ \delta^0_{~\mu} \epsilon(x) \\
\delta_{\eta} A_{\mu}(x) & = &\left\{A_{\mu}(x), \int d^3 x'
\eta(x') \partial_i \pi^i(x')\right\} ~~=~~ - \delta^i_{~\mu} \partial_i
\eta(x) \\
\delta \pi(x) & = & 0
\end{eqnarray}

Clearly, $\delta_{\epsilon}A_0(x) = \epsilon(x), ~\delta_{\eta}A_i(x) = -
\partial \eta (x)$. {\em Choosing} $\epsilon(x) := - \partial_0
\eta(x)$, reveals that the canonical transformations generated by the
constraints are nothing but the usual {\em gauge transformations}.

Let us summarize.
\begin{enumerate}
\item Maxwell equations can be written in a manifestly relativistic
form.

\item These can be derived from relativistic action principle treating
$A_{\mu}(t, x^i)$ as generalized coordinates. In the Lagrangian
formulation, the `matrix of second derivatives', 
\begin{equation}
\frac{\delta^2 {\cal L}}{\delta (\partial_0 A_{\mu}(x)) \delta (\partial_0
A_{\nu}(x'))} ~:=~ {\cal M}^{\mu\nu}(x, x') ~=~ \left(- \eta^{\mu\nu} +
\eta^{0\mu} \eta^{0\nu} \right) \delta^3(x - x')\ ,
\end{equation}
is {\em non-invertible}.

\item In the Hamiltonian formulation, there are {\em constraints}. These
generate canonical transformations which are the familiar {\em gauge 
transformations}. 

\item The canonical Hamiltonian has the form which contains the
constraints with arbitrary functions of time as coefficients.

\item We also know that while manifest special relativistic formulation
requires us to use 4 components of the vector potential as basic
configuration space variables (i.e. 4 degrees of freedom per point),
physically there are only 2 degrees of freedom (the two polarizations).
Thus the constraints inferred above have something to do with
identifying {\em physical degrees of freedom}.
\end{enumerate}

We will keep these in mind and try to abstract these features in a
general framework. On the one hand we will {\em simplify} by restricting
to {\em finitely many} degrees of freedom and at the same time
generalize to arbitrarily complicated systems.

\chapter{Constrained Lagrangian and Hamiltonian Systems}

Consider a dynamical system with finitely many degrees of freedom,
described in a Lagrangian framework. Let $Q$ be an $n$-dimensional
manifold with local coordinates $q^i$ serving as generalized
coordinates. Let $L(q, \dot{q})$ be the Lagrangian function of the
dynamical system and $S[q(t)] := \int dt L(q, \dot{q})$ being the
action. The Variational Principle leads to the Euler-Lagrange equations
of motion,
\begin{eqnarray}
\frac{d}{dt} \left(\frac{\partial L}{\partial \dot{q}^i}\right) & = &
\frac{\partial L}{\partial q^i} \hspace{4.8cm} i = 1, 2, \cdots, n\ , \\
\left(\frac{\partial^2 L}{\partial \dot{q}^i \partial \dot{q}^j}\right)
\ddot{q}^j & := & M_{ij}(q, \dot{q}) ~ \ddot{q}^j ~ = ~ 
\frac{\partial L}{\partial q^i} - \frac{\partial^2 L}{\partial \dot{q}^i
\partial q^j} \dot{q}^j ~ := ~ Q_i(q, \dot{q}) \ .
\end{eqnarray}

Usually, the matrix $M_{ij}$ is invertible which allows us to solve for
the accelerations, $\ddot{q}^i = (M^{-1})^{ij} Q_j$. This implies that
given position and velocity at an instance, one can always determine the
dynamical trajectory at other instances.

If on the other hand the matrix $M_{ij}$ is {\em not} invertible, then
let $1 \le r \le n$, denote its rank. Then there exist $(n - r)$
independent $h_a^i$, vectors satisfying $M h = 0$ and we cannot solve
for {\em all} the accelerations {\em uniquely}. The general solution for
the accelerations will be, $\ddot{q}^i = \underline{\ddot{q}^i} +
\sum_{a = 1}^{n - r} \alpha_a h^i_a$ and the first term is the solution
of the inhomogeneous equation. Furthermore, $Mh = 0$ implies $h^TM = 0 =
h_a^i Q_i(q, \dot{q}) = 0$ for $a = 1, \cdots (n -r)$. Thus on the one
hand the equations of motion are {\em not} uniquely specified and on the
other hand there are $(n - r)$ relations among the $2n$ positions and
velocities. If these relations are {\em not} satisfied identically, then
in particular, they correspond to {\em restrictions on the initial data,
$q^i(0), \dot{q}^i(0)$}.

As an example, consider a system with a one dimensional
configuration space with coordinate $q$ and with a Lagrangian, $L = f(q)
\dot{q} - V(q)$. Clearly, $M = 0$ and $Q = -V'(q)$. The acceleration is
given by $\ddot{q} = \alpha h$ and $h Q = 0 \Rightarrow V'(q) = 0$. If
$V = 0$ then the equation is satisfied identically while for non-zero
$V$, the position must be at an extremum of $V$ (which may not exist!).

Let us see what implications are there for a corresponding Hamiltonian
framework. First observation is that for a singular Lagrangian, the
definition of conjugate momentum $p_i := \frac{\partial L}{\partial
\dot{q}^i}$ {\em cannot} be inverted to solve for the velocities in
terms of the momenta. This is essentially the {\em inverse
function theorem} namely, $y^i = f^i(x)$ can be inverted to get $x^i =
g^i(y)$ provided $\frac{\partial f^i}{\partial x^j}$ is an
invertible matrix. (Strictly speaking, the $n$ momenta are $n$ functions
of the $2n$ positions and velocities and hence one should be invoking the
{\em implicit function theorem}, such fine prints are ignored here.)

Let us define the Hamiltonian by $H := p_i \dot{q}^i - L(q, \dot{q})$
and consider its variation,
\begin{eqnarray}
\delta H & = & \dot{q}^i \delta p_i + \left(p_i - \frac{\partial L}{\partial
\dot{q}^i}\right) \delta \dot{q}^i - \frac{\partial L}{\partial q^i} \delta
q^i
\end{eqnarray}
If we treat the $q, \dot{q}, p$ all as independent variables, then the
`Hamiltonian' is clearly a function of all of these. Let us {\em use}
the definition of momenta (i.e. restrict to a $2n$ dimensional
sub-manifold of the $3n$ dimensional space). On this sub-manifold, the
middle term vanishes and the Hamiltonian becomes a function {\em only}
of positions and momenta in the sense that it varies only when $q^i,
p_i$ are varied.  This is {\em independent} of whether velocities can be
solved for in terms of momenta and positions. The next step in the usual
case of non-singular Lagrangians is when one infers the Hamilton's
equations of motion as,
\begin{equation}
\dot{q}^i ~ = ~ \frac{\partial H}{\partial p_i} ~~~,~~~
\dot{p}_i ~ = ~ - \frac{\partial H}{\partial q^i} \ .
\end{equation}
This assumes that (a) the variations $\delta p_i, \delta q^i$ are {\em
independent} and (b) Hamiltonian is a function {\em only} of $p_i, q^i$.
Both these statements fail for singular Lagrangians.

To see this, notice that variation of the Hamiltonian being given in
terms of variations of momenta and positions, depends on evaluating the
variation of the Hamiltonian on the sub-manifold defined by $p_i =
\frac{\partial L}{\partial \dot{q}^i}$. If the variations are also to
respect this condition, we must have 
\begin{eqnarray}
\delta p_i & = & M_{ij} \delta \dot{q}^j  + \frac{\partial^2 L}{\partial
\dot{q}^i \partial q^j} \delta q^j \\
h^i_a \delta p_i & = & 0 + h^i_a \frac{\partial^2 L}{\partial
\dot{q}^i \partial q^j} \delta q^j 
\end{eqnarray}
which immediately shows that the variations $\delta p_i, \delta q^i$ are
{\em not independent}. Consequently, one cannot deduce the Hamilton's
equations of motion. We have already noted that for singular
Lagrangians, the velocities cannot be eliminated in favour of momenta
and positions. The variations being not independent means that there are
relations among the phase space variables $p_i, q^i$, namely, $\phi_a(q, p) =
0$.

Let us summarize. 

A Lagrangian system is said to be {\em singular} if the matrix $M_{ij}$
of second derivatives of the Lagrangian with respect to the velocities,
is singular. This has the consequences that (a) the accelerations are
not determined uniquely i.e. contain arbitrary functions of time and (b)
it could imply relations among velocities and positions which may not
even be consistent. It is possible to isolate the accelerations which
are determined uniquely. However we will carry out such analysis in the
context of a Hamiltonian framework. The property of being a singular
Lagrangian is {\em independent} of any choice of local coordinates.
Generically, singular Lagrangian systems are also called {\em
constrained systems} \cite{Sudarshan}.

In a Hamiltonian formulation obtained from a Lagrangian, the singular
nature of the Lagrangian manifests as certain relations among the phase
space variables. We will therefore define a {\em constrained Hamiltonian
System} to be a system with a $2 n$ dimensional phase space manifold
$\Gamma$ with local coordinates $(q^i, p_i) \leftrightarrow
\omega^{\mu}$, on which is given a Hamiltonian function, $H_0(q^i, p_i) :=
H_0(\omega)$ {\em together with a set of relations $\phi^a(\omega) = 0, a
= 1, \cdots k (< 2 n)$}. These relations are referred to as {\em Primary
Constraints}. By definition, the $k$ constraints are {\em functionally
independent} i.e. the $k$ differentials, $d \phi^a(\omega)$ (or the $k$
vectors, $\partial_{\mu} \phi^a$) are linearly independent. The
sub-manifold defined by $\phi^a = 0$ will be denoted by $\Sigma$ and is
called the {\em constraint surface}. It is a $2 n - k$ dimensional
sub-manifold of $\Gamma$. Note that a constraint surface is {\em not a
phase space} in general, i.e. does not have a symplectic form (eg when
$k$ is odd).

We will focus entirely on the constrained Hamiltonian systems and
analyze various possibilities of types of such systems. The aim will be
to have {\em a procedure for obtaining equations of motion} in the {\em
Hamiltonian form} paying attention to the constraints. This means that
we want to have a variational principle for paths in $\Gamma$, which
will lead to `Hamilton's equations of motion' for some suitable
Hamiltonian function and such that the possible dynamical trajectories
(in $\Gamma$) either remain confined to the constraint surface,
$\Sigma$ or never intersect it. In effect, we can continue to work in
the given phase space and use a {\em new} Hamiltonian function so that
dynamics effectively respects the constraints.  This is achieved by using
the method of Lagrange multipliers. 

Introduce $k$ Lagrange multipliers, $\lambda_a$ and define a new
Hamiltonian function $H := H_0 + \lambda_a \phi^a$, which matches with
the given Hamiltonian
$H_0$ {\em on the constraint surface}, $\Sigma$. Defining an action,
\begin{equation}
S[\omega(t), \lambda(t)] ~ := ~ \int dt \left[ 
\frac{1}{2}\omega^{\mu} \Omega_{\mu\nu} \dot{\omega}^{\nu} - 
H_0(\omega) - \lambda_a(t) \phi^a(\omega)\right] \ ,
\end{equation}
and invoking its stationarity, $\delta S = 0$, leads to the equations of
motion,
\begin{equation}
\frac{d \omega^{\mu}}{d t} ~=~ \Omega^{\mu\nu} \left(
\frac{\partial H_0}{\partial \omega^{\nu}} +
\lambda_a \frac{\partial \phi^a}{\partial \omega^{\nu}}\right) 
~~~\mathrm{and}~~~ \phi^a (\omega) = 0 ~~,~~ a = 1, \cdots, k\ .
\end{equation}
Thus we obtain equations of motion in a Hamiltonian form and also the
constraint equations. We have succeeded in having a new Hamiltonian dynamics
defined for trajectories in $\Gamma$. We have now to ensure that the
trajectories are such that if an initial point is on the constrained
surface, then the whole trajectory also remain on the constrained
surface. If this property can be ensured, then it also follows that no
trajectory can enter and leave $\Sigma$, since the equations are first order. 

Observe that if the value of $\phi^a$ for any given $a$ is preserved
under evolution (i.e. $\dot{\phi}^a = 0$ along a trajectory), then the
trajectory is confined to the $2 n - 1$ dimensional hyper-surface
$\phi^a = $constant.  However we only need the trajectory to be confined
to $\Sigma$, so the value of each constraint need {\em not} be preserved
{\em exactly}.  Therefore we need {\em not} have $\dot{\phi^a} = 0$
along {\em all} trajectories but {\em only along those trajectories
which lie in $\Sigma$}. This is ensured by requiring that the Poisson
bracket of the constraints with the Hamiltonian $H$ be {\em weakly
zero}. Since Poisson brackets can be evaluated at any point of the phase
space, we can evaluate these at points on the constraint surface and
{\em weak} equality/equations refer to Poisson brackets being evaluated
at points of $\Sigma$ and are denoted by `$\approx$' \cite{Dirac}.
Strong equations/equalities are valid in a {\em neighbourhood of
$\Sigma$} and are denoted by the usual `$=$'. In particular a strongly
vanishing function vanishes weakly and so do {\em all its partial
derivatives}.
\begin{equation} \label{ConsistencyCond} \frac{d \phi^a}{d t} ~=~
\{\phi^a, H_0\} + \lambda_b \{\phi^a, \phi^b\} ~\approx~ 0
\end{equation} These $k$ conditions ensure that a trajectory beginning
on the constrained surface remains on the constraint surface and our
goal is reached {\em provided} (\ref{ConsistencyCond}) {\em holds on
$\Sigma$}.

There are several possibilities now \cite{Dirac}.

{\em If } the $k \times k$ matrix of Poisson brackets of the constraints
is {\em non-vanishing}, then the consistency conditions can be viewed as
a matrix equation at each point on $\Sigma$ {\em for the Lagrange
multipliers $\lambda_a$}. Since this matrix is antisymmetric, it is {\em
non-singular} only if when $k$ is even. In this case, all Lagrange
multipliers are determined and we do have a Hamiltonian dynamics whose
trajectories either lie on the constraint surface or never intersect it.
Generically however the matrix is singular. As seen in the context of
singular Lagrangian, this means that (a) some multipliers are
necessarily undetermined and (b) some linear combinations of $\{\phi^a,
H_0\}$ must vanish on $\Sigma$. Again we have several possibilities.
Either (i) $\{\phi^a,H_0\} \approx 0, \forall a$ and {\em all} linear
combinations are weakly zero, or (ii) the linear combinations vanish
weakly {\em provided some further functions vanish} in which case we
refer to these as {\em secondary constraints}, or (iii) there are {\em
no} points of $\Sigma$ at which the linear combinations vanish in which
case we say that the Hamiltonian system is {\em inconsistent}. 

It could also happen that the matrix is zero on the constraint surface.
This could happen for instance, if $\{\phi^a, \phi^b\} =
C^{ab}_{~~c}\phi^c$. (Recall the Maxwell example). In such a case, there
is no equation for the Lagrange multipliers and {\em all} Lagrange
multipliers are undetermined. This case can be thought of as a special
case of rank of the matrix being zero. 

If we encounter the inconsistent case, we throw away our formulation of 
the system and start all over again. In the case (i), we have reached
our goal but have to live with some undetermined Lagrange multipliers
(and hence evolution). This will turn out to be the most interesting
case. In the case (ii), we have to now demand that the Poisson bracket
of secondary constraints with the Hamiltonian $H$ must vanish on
$\Sigma$. Once again we will encounter similar cases as above and we
have to repeat the analysis -- we will either satisfy the conditions
identically with some Lagrange multipliers determined or encounter {\em
tertiary constraints} or encounter inconsistency. Since the total number
of constraints cannot be larger that $2n$ (else no initial condition
will be left!), the process must terminate. Barring inconsistent
systems, we will eventually end up with some Lagrange multipliers being
determined, some undetermined and with possibly additional constraints
$\chi_{A} \approx 0, A = 1, \cdots, l$.

Note: We began by requiring the trajectories to be confined to $\Sigma$
and found as a consistency requirement that the goal cannot be attained
{\em for all points of $\Sigma$}. We need to restrict further to a
sub-manifold $\Sigma' \subset \Sigma$, due to the secondary constraints.
Thus, {\em the dynamics defined by $H$ on $\Gamma$, will correspond to a
constrained dynamics relative to $\Sigma'$ defined by all constraints
being weakly zero}. The dimension of $\Sigma'$ is of course $2 n - k -
l$. Notice that even if we began by requiring consistency condition to
hold on $\Sigma$, extending it to hold on $\Sigma'$, does not contradict
the previous condition since $\Sigma'$ is a sub-manifold of $\Sigma$.
The weak/strong equation now refer to $\Sigma'$.

To summarize: Beginning with a phase space $\Gamma$, a Hamiltonian $H_0$
and a set of primary constraints $\phi^a$, we {\em can} construct a
Hamiltonian dynamics on $\Gamma$ such that its trajectories are either
confined to the constraint surface $\Sigma$ or avoid it. The
construction reveals the possibility of further constraints as well as
the dynamics being {\em not completely determined}.  In the next
lecture, we will consider a suitable classification of constraints and
obtain a corresponding classification of constrained Hamiltonian
systems.

\newpage
Here are some elementary examples.
\begin{enumerate}

\item $H_0 = \Case{p^2}{2 m}, ~ \phi(q) = q - q_0~~:~~$ $\{\phi, H\} ~=~
\frac{p}{m} \approx 0 ~$, $p \approx 0$, is a secondary constraint;

\item $H_0 = \Case{p^2}{2 m}, ~ \phi(q) = q^2 ~~:~~$ $\{\phi, H\} ~=~ \frac{2
q p}{m} \approx 0 ~~~$condition holds; 

\item $H_0 = a p, \ a \ne 0, ~ \phi(q) = q ~~:~~$ $ \{\phi, H\} ~=~ a
\approx 0 ~~~$inconsistency.

\end{enumerate}

\chapter{Dirac-Bergmann theory of Constrained Hamiltonian Systems}

Our consistent constrained Hamiltonian system is specified by,
\begin{equation}
H~=~ H_0 + \sum_{a = 1}^k \lambda_a \phi^a~~,~~\phi^a \approx 0~~,~~
\chi^A \approx 0, ~~ A = 1, \cdots l~,~~ k + l < 2 n\ .
\end{equation}
where `$\approx$' means evaluation on $\Sigma$ defined by the primary
constraints ($\phi^a \approx 0$) and the secondary constraints ($\chi^A
\approx 0$). The constraints also have to satisfy,
\begin{equation} \label{ConsCond}
\{\phi^a, H_0\} + \{\phi^a, \phi^b\} \lambda_b ~\approx~ 0~~~,~~~
\{\chi^A, H_0\} + \{\chi^A, \phi^b\} \lambda_b ~\approx~ 0 \ .
\end{equation}
Thus we have $k + l$ equations for $k$ Lagrange multipliers and the
system is naively, over-determined. 

Let rank of the $(k + l) \times k$ matrix of Poisson brackets of the
constraints be $K \le k$. This means that $K$ is the maximum number of
linearly independent rows and columns of the matrix. Thus there exist $k
- K$ independent relations among the $k$ columns of the matrix i.e.
$\exists$ $\xi^{(\alpha)}_b, \ \alpha = 1, \cdots, k - K$ numbers such
that
\begin{equation}
\{\phi^a, \phi^b\} \xi^{(\alpha)}_b \approx 0 ~~,~~
\{\chi^A, \phi^b\} \xi^{(\alpha)}_b \approx 0 \ .
\end{equation}
Now define two sets of linear combinations of the primary constraints
namely,
\begin{equation}
\tilde{\phi}^{\alpha} ~:=~ \xi^{(\alpha)}_a \phi^a ~~~,~~~
\tilde{\phi}^{\alpha'} ~:=~ \eta^{(\alpha')}_a \phi^a ~~~,~~~ \alpha = 1,
\cdots, k - K~~~ \mbox{and}~~~\alpha' = 1, \cdots, K \ .
\end{equation}
The new set of $K$ vectors $\eta^{(\alpha')}_a$ are so chosen that the
the set of constraints $\tilde{\phi}^{\alpha}, \tilde{\phi}^{\alpha'}$
are functionally independent. Now it is clear that
\begin{equation}
\{\phi^a, \tilde{\phi}^{\alpha}\} \approx 0 ~~,~~
\{\chi^A, \tilde{\phi}^{\alpha}\} \approx 0 \ .
\end{equation}
Thus the $k - K$ new combinations, $\tilde{\phi}^{\alpha}$ of primary
constraints have a weakly vanishing Poisson bracket with {\em all} the
constraints. Such constraints are termed {\em first class constraints}.
Constraints which do not have this property are termed {\em second class
constraints}. With the help of the $\xi$'s we have regrouped the primary
constraints into primary first class and primary second class
constraints. 

We would like to do the same for secondary constraints. Observe that
linear combination of second class constraints will again satisfy the
consistency conditions and to such combinations could be added any
combination of primary constraints without affecting these equations. To
maintain the functional independence of the secondary constraints we
need to ensure that the linear combinations are also functionally
independent. Thus, consider the combinations,
\begin{equation}
\tilde{\chi}^{A} ~:=~ S^{A}_{~B}\chi^B + S^A_{~\alpha}\phi^{\alpha} +
S^A_{~\alpha'} \phi^{\alpha'}
\end{equation}
with $S^{\cdot}_{~\cdot}$ so chosen that $S^A_{~B}$ is a non-singular matrix
and $\tilde{\chi}^A$ have weakly vanishing Poisson bracket with {\em
all} constraints for a {\em maximal number} of values of $A$. Let this
number be $L \le l$. Thus, we divide the combinations $\tilde{\chi}$
into the first class combinations, $\tilde{\chi}^{\sigma}, \sigma = 1,
\cdots, L$ and the second class combinations $\tilde{\chi}^{\sigma'}, 
\sigma' = 1, \cdots, l - L$.

The result of these manipulations is that (a) we can write $\lambda_a
\phi^a = {\tilde\lambda}_{\alpha}\tilde{\phi}^{\alpha} +
{\tilde\lambda}_{\alpha'} \tilde{\phi}^{\alpha'}$ and (b) the
consistency conditions can be simplified as,
\begin{eqnarray}
\{\tilde{\phi}^{\alpha}, H_0\} \approx 0 &,& 
\{\tilde{\chi}^{\sigma}, H_0\} \approx 0 \label{ConsCondFirst}\\
\{\tilde{\phi}^{\alpha'}, H_0\} + \tilde{\lambda}_{\beta'}
\{\tilde{\phi}^{\alpha'}, \tilde{\phi}^{\beta'}\} \approx 0 &,& 
\{\tilde{\chi}^{\sigma'}, H_0\} + \tilde{\lambda}_{\beta'}
\{\tilde{\chi}^{\sigma'}, \tilde{\phi}^{\beta'}\} \approx 0 \ .
\label{ConsCondSecond}
\end{eqnarray}
The first set of equations involve only the $k - K + L$ first class
constraints and no Lagrange multipliers while the second set of $K + l -
L$ equations involve only the $K$ Lagrange multipliers,
$\tilde{\lambda}_{\alpha'}$.  The $k - K$ Lagrange multipliers,
$\tilde{\lambda}_{\alpha}$ have dropped out of the equations and will
remain {\em undetermined}. Once again we have more equations than
unknown, but because of the separation into first and second class
constraints, we are now guaranteed that the $(K + l - L) \times K$
matrix of Poisson brackets of $\tilde{\phi}^{\alpha'}$ and
$\tilde{\chi}^{\sigma'}$ with $\tilde{\phi}^{\beta'}$ has the maximal
rank $K$. For, if it did not, there would exist further linear
combinations which will weakly Poisson commute with all constraints and
by construction we have obtained the maximum number of first class
constraints. We will now solve for the Lagrange multipliers
$\tilde{\lambda}_{\alpha'}$ explicitly.

Define the matrix $\Delta$ of Poisson brackets of the second class
constraints as,
\begin{eqnarray}
\Delta & := & \left( 
\begin{array}{cc}
\{\tilde{\phi}^{\alpha'}, \tilde{\phi}^{\beta'}\} &
\{\tilde{\phi}^{\alpha'}, \tilde{\chi}^{\sigma'}\} \\
\{\tilde{\chi}^{\rho'}, \tilde{\phi}^{\beta'}\} &
\{\tilde{\chi}^{\rho'}, \tilde{\chi}^{\sigma'}\}
\end{array}
\right) 
~~ := ~~ \left(
\begin{array}{cc}
\Delta^{\alpha' \beta'} & \Delta^{\alpha' \sigma'} \\
\Delta^{\rho' \beta'} & \Delta^{\rho' \sigma'} 
\end{array}
\right) 
\end{eqnarray}
This is an antisymmetric square matrix of order $K + l - L$. This
{\em must be non-singular}. For, if it is singular, there will exist
non-trivial linear combination of the second constraints
$\tilde{\phi}^{\alpha'}, \tilde{\chi}^{\sigma'}$, which will weakly
Poisson commute with all the second class constraints (and it
automatically commutes with the first class constraints), implying that
we have additional first class constraint. The non-singularity also
requires that the total number of second class constraints, $K + l -
L$, must be an {\em even integer}. Let its (weak) inverse be the matrix
$C$,
\begin{eqnarray}
C & := & \left( 
\begin{array}{cc}
C_{\alpha' \beta'} & C_{\alpha' \sigma'} \\
C_{\rho' \beta'} & C_{\rho' \sigma'} 
\end{array}
\right) 
\end{eqnarray}
The equation $C \Delta \approx \mathbb{1}$ translates into,
\begin{eqnarray}
C_{\alpha'\beta'}\Delta^{\beta'\gamma'} +
C_{\alpha'\sigma'}\Delta^{\sigma'\gamma'} 
& = & \delta^{~~\gamma'}_{\alpha'} \label{First}\\
C_{\rho'\beta'}\Delta^{\beta'\tau'} +
C_{\rho'\sigma'}\Delta^{\sigma'\tau'} 
& = & \delta^{~~\tau'}_{\rho'} \\
C_{\alpha'\beta'}\Delta^{\beta'\tau'} +
C_{\alpha'\sigma'}\Delta^{\sigma'\tau'} & = & 0 \\
C_{\rho'\beta'}\Delta^{\beta'\gamma'} +
C_{\rho'\sigma'}\Delta^{\sigma'\gamma'} & = & 0 \label{Fourth}
\end{eqnarray}

The equations for the Lagrange multipliers become,
\begin{eqnarray}
\{\tilde{\phi}^{\beta'}, H_0\} + 
\Delta^{\beta'\gamma'}\tilde{\lambda}_{\gamma'} \approx 0 & , & 
\{\tilde{\chi}^{\sigma'}, H_0\} + 
\Delta^{\sigma'\gamma'}\tilde{\lambda}_{\gamma'} \approx 0
\end{eqnarray}
Multiply the first one by $C_{\alpha'\beta'}$, second one by
$C_{\alpha'\sigma'}$, add the two and use (\ref{First}) to solve for
$\tilde{\lambda}_{\alpha'}$. One gets, 
\begin{equation} \label{LambdaSoln}
\tilde{\lambda}_{\alpha'} ~\approx~ 
- C_{\alpha'\beta'} \{\tilde{\phi}^{\beta'}, H_0\} 
- C_{\alpha'\sigma'} \{\tilde{\chi}^{\sigma'}, H_0\}
\end{equation}
Similar manipulation using (\ref{Fourth}) equation leads to,
\begin{equation} \label{ForSymmetrization}
C_{\rho'\beta'} \{\tilde{\phi}^{\beta'}, H_0\} + 
C_{\rho'\sigma'} \{\tilde{\chi}^{\sigma'}, H_0\} ~\approx ~ 0
\end{equation}

We can now write the Hamiltonian explicitly using the solution
(\ref{LambdaSoln}) and express the evolution of any function on the
phase space as its Poisson bracket with $H$. We will write it in a more
convenient form.
\begin{eqnarray} \label{Evolution}
\frac{d }{d t} f(\omega(t)) & \approx & \{f, H_0\} + \lambda_{\alpha}
\{f, \tilde{\phi}^{\alpha}\} \nonumber \\
& & 
- \{f, \tilde{\phi}^{\alpha'}\} C_{\alpha'\beta'} \{\tilde{\phi}^{\beta'}, H_0\}
- \{f, \tilde{\phi}^{\alpha'}\} C_{\alpha'\sigma'} \{\tilde{\chi}^{\sigma'},
H_0\} \\
& & 
- \{f, \tilde{\chi}^{\rho'}\} C_{\rho'\beta'} \{\tilde{\phi}^{\beta'}, H_0\} 
- \{f, \tilde{\chi}^{\rho'}\} C_{\rho'\sigma'} \{\tilde{\chi}^{\sigma'}, H_0\}
\nonumber
\end{eqnarray}
The last line, which is weakly zero due to (\ref{ForSymmetrization}),
has been added to get a more symmetric final expression.

Now let us denote all the second class constraints,
$(\tilde{\phi}^{\alpha'}, \tilde{\chi}^{\rho'})$ by $\xi^m, m = 1,
\cdots, K + l - L$. Then, the nonsingular matrix $\Delta$ is just the
matrix $\Delta^{mn} = \{\xi^{m}, \xi^n\}$, $C_{mn}$ is its weak inverse
as before and the last group of four terms in (\ref{Evolution}) are
conveniently expressed as $- \{f, \xi^m\}C_{mn}\{\xi^n, H_0\}$ so that
finally we obtain,
\begin{eqnarray}\label{FinalEvolution}
\frac{d }{d t} f( \omega(t) ) & \approx & \{f, H_0\} + \sum_{\alpha}
\lambda_{\alpha} \{f, \phi^{\alpha}\}  
- \{f, \xi^m\}(\Delta^{-1})_{mn}\{\xi^n, H_0\} ~, \\
& :\approx & \{f, H_0 + \sum_{\alpha}\lambda_{\alpha} 
\phi^{\alpha}\}^* \hspace{4.1cm}\mbox {where,} \nonumber \\ 
\Delta_{mn} & := & \{\xi^m, \xi^n\} \hspace{6.2cm}\mbox{and} \\
\{f, g\}^* & := & \{f, g\} - \{f, \xi^m\}(\Delta^{-1})_{mn}\{\xi^n,
g\} \hspace{2.0cm}\mbox{(Dirac Bracket)}\ . \label{DiracBracketDefn}
\end{eqnarray}
In this final expression, we have removed the $~\tilde{}~$, the primary
first class constraints are denoted by $\phi^{\alpha}$ while all second
class constraints are denoted by $\xi^m$.

Several remarks are in order.
\begin{enumerate}
\item The first step in the analysis of constrained systems was to
obtain the full set of constraints starting with a given Hamiltonian
$H_0$ and a set of primary constraints defining the constraint surface
$\Sigma \subset \Gamma$. In order to ensure that we get the final form
of evolution equations to be a Hamiltonian form, we used a modified
Hamiltonian $H$ and thought of the system as thought it were {\em
un-constrained} in the sense the variations of the phase space
coordinates were {\em independent}. To make contact with the constrained
nature of the system, we required that the un-constrained dynamics be
such that its trajectories either lie in $\Sigma$ or never intersect it.
This lead us to discovering possible, additional constraints. Note that
restrictions on the trajectories was with reference to the sub-manifold
$\Sigma$ and hence only the primary constraints played a role in
subsequent analysis i.e. we did {\em not} add to the Hamiltonian, terms
corresponding to the secondary constraints. The secondary constraints
however do reveal that the required segregation of trajectories holds
only with respect $\Sigma' \subset \Sigma$, defined by vanishing of all
constraints. 

\item The next step was essentially an exercise in linear algebra. We
did this to solve explicitly for those Lagrange multipliers which could
be solved for. This was facilitated by regrouping the set of all
constraints into first class and second class constraints. The final
result reveals that evolution could be {\em arbitrary} if there are {\em
primary, first class constraints} due to the undetermined
$\lambda_{\alpha}$. The most compact expression for the Hamiltonian
evolution was obtained using the {\em Dirac brackets}.

\item We now define {\em first class variables} as those functions on
$\Gamma$ whose Poisson brackets with {\em all} constraints are weakly
zero. By the consistency condition (\ref{ConsCond}), the Hamiltonian $H$
is a first class variable and of course so are the first class
constraints. The Hamiltonian $H_0$ may or may not be a first class
variable. It's Poisson bracket with first class constraints is of course
weakly zero from (\ref{ConsCondFirst}). It is easy to check that sums
and products of first class variables is again first class and so are the
Poisson brackets of first class variables 
\footnote{Functions on the phase space of an unconstrained system are
generally called {\em observables} while in the context of a gauge
system, the first class observables are all called as {\em Dirac
observables}.}.

\item As noted already, in the presence of primary first class
constraints, evolution equation for a generic function $f$, contains the
arbitrary Lagrange multipliers, $\lambda_{\alpha}$. From
(\ref{FinalEvolution}), it follows that {\em evolution of first class
variables is independent of $\lambda_{\alpha}$ and is entirely governed
by $H_0$}. This justifies why first class variables are singled out.

\item The Dirac brackets have been introduced as a convenient compact
notation. However it has many interesting properties, namely,
\begin{enumerate}
\item $\{f, g + h\}^* ~\approx~ \{f, g\}^* + \{f, h\}^*$ (addition);
\item $\{f, \mu g\}^* ~\approx~ \mu \{f, g\}^* $ (scalar multiplication);
\item $\{f, g h\}^* ~\approx~ \{f, g\}^* h + g \{f, h\}^*$ (Leibniz);
\item $\{f, g\}^* ~\approx~ - \{g, f\}^*$ (antisymmetry);
\item $\{f, \{g, h\}^*\}^*$ + cyclic $~\approx~ 0$ (Jacobi identity).
\end{enumerate}
Thus it has all the properties of the usual Poisson bracket. In
addition, one has
\begin{enumerate}
\item $\{f, g\}^* ~\approx~ \{f, g\}$ for any first class $f$ and 
$\forall g$;
\item $\{\xi^i, g\}^* ~\approx~ 0, \forall g$ and any second class
constraint $\xi^i$.
\end{enumerate}
Recall that weak equations involving Poisson brackets mean that the
Poisson brackets are {\em first computed in a neighbourhood of
$\Sigma$} and {\em then evaluated on $\Sigma$}. This rule is necessary
since a weakly vanishing function need {\em not} have a weakly vanishing
Poisson bracket (since some of the partial derivatives `off' $\Sigma$
may not be zero). This applies to the second class constraints as well.
However, Dirac bracket of a second class constraint with {\em any}
function is weakly zero. Therefore, if we use Dirac brackets for writing
the equations of motion (as shown in (\ref{FinalEvolution})), then we
can set the second class constraints to be zero {\em before} computing
the Dirac brackets. This is equivalent to reducing the phase space
dimension from $2 n$ to ($2 n$ - the number of second class
constraints). Thus, {\em second class constraints correspond to
redundant degrees of freedom which can be ignored by setting them to
zero}. 

\item We could now focus on systems that do {\em not} have any second
class constraints either a priori or after eliminating them via the
Dirac bracket procedure. One is effectively left systems with only first
class constraints. As noted earlier, the evolution of a dynamical
variable in such systems is in general, arbitrary and only variables
whose evolution is {\em not} arbitrary are the first class variables for
whom the Dirac brackets are same as the Poisson brackets.

Note that it could happen that there are {\em no} first class constraints
left. In such a case, we have just the usual types of systems. Thus, the
net conclusion of the analysis is: 

Generically, consistent Hamiltonian systems are systems with first class
constraints with the special case of no first class constraints.
Hamiltonian systems with at least one first class constraint are termed
{\em gauge theories}.  We will now focus on these exclusively.
\end{enumerate}


Let $\Gamma$ be a $ 2n$ dimensional phase on which is given a first
class Hamiltonian function $H_0$ and a set of of $k < 2n$ first class
constraints, $\phi^a$ whose vanishing defines the constraint surface
$\Sigma$. The total Hamiltonian governing time evolutions is given by $H
:= H_0 + \sum_a \lambda_a \phi^a$. By virtue of being first class, we
have the following relations:
\begin{equation}
\{\phi^a, \phi^b\} ~\approx~ 0 \leftrightarrow 
\{\phi^a, \phi^b\} ~=~ C^{ab}_{~~c}(\omega) \phi^c ~~\mbox{and}~~
\{H_0, \phi^a\} ~\approx~ 0 \leftrightarrow 
\{H_0, \phi^a\} ~=~ D^{a}_{~b}(\omega) \phi^b \ .
\end{equation}
The evolution equation for any function $f$ on $\Gamma$ is given by,
\begin{equation}
\frac{d}{d t} f(\omega(t)) ~\approx~ \{ f(\omega), H_0 + \sum_a \lambda_a
\phi^a(\omega)\}|_{\omega = \omega(t)}
\end{equation}

Now observe that (a) if we make an arbitrary {\em diffeomorphism} i.e. a
mapping of the manifold $\Gamma$ on to itself preserving differential
structure, the manifold is unchanged (by definition). However, the
symplectic form would change in general; (b) if we specialize the
diffeomorphisms to those which preserve the symplectic form, then the
restricted diffeomorphisms are the familiar {\em canonical
transformations}. All such (continuously connected to identity)
transformations can be generated by arbitrary functions on $\Gamma$, by
the rule: $\delta_{\epsilon g} \omega^{\mu} := \epsilon
\Omega^{\mu\nu}\partial_{\nu} g(\omega)$. All of these however do {\em
not} preserve the constraint surface; (c) To preserve the constraint
surface, the function must preserve the constraints defining the surface
i.e. must be a first class function. Thus, {\em all} first class
functions and in particular the first class constraints, do preserve
$\Sigma$. While constraints preserve the Hamiltonian, other first class
functions need not. Those first class functions which do preserve the
Hamiltonian as well are said to generate {\em symmetry transformations}.
By contrast, the transformations generated by the first class
constraints are distinguished as {\em gauge transformations}.

Thus, the diffeomorphisms generated by first class constraints, preserve
the entire structure of the constrained Hamiltonian system (i.e.
manifold, symplectic structure, constraint surface and the Hamiltonian)
and are termed {\em gauge transformations}. First class functions (which
are not constraints) are automatically invariant under these
transformations. Provided they Poisson commute with the Hamiltonian,
they generate {\em symmetry transformations}. This distinction among the
set of all first class function, comes about for the following reason.

Recall that $\lambda_a$ are undetermined, arbitrary functions that
appear in the equations of motions. Therefore, beginning from any
initial condition $\hat{\omega} \in \Sigma$, the actual trajectories
will depend on the {\em choice} made for the $\lambda_a$'s. {\em If} we
identified points on $\Sigma$ as representing physical states of the
system, we would loose determinism -- a given state does {\em not}
uniquely determine the future state. We need to identify physical states
of the system differently. 

Consider infinitesimal evolutions for two different choices of
$\lambda_a$'s. We will have,
\begin{equation}
\omega' (\delta t) ~=~ \hat{\omega} + \delta t \{\omega, H(\lambda', \omega)\}
|_{\hat{\omega}} ~~,~~
\omega (\delta t) ~=~ \hat{\omega} + \delta t \{\omega, H(\lambda, \omega)\}
|_{\hat{\omega}} \ ,
\end{equation}
which implies that,
\begin{equation}
\delta \omega(\delta t) ~=~ \delta t \{\omega, \sum_a \delta \lambda_a
\phi^a\}|_{\hat{\omega}} ~~=~~\sum_a (\delta t \delta \lambda_a)
\{\omega, \phi^a\}|_{\hat{\omega}} \ ,
\end{equation}
which is nothing but the infinitesimal transformation generated by the
first class constraints! 

Thus, if we arbitrarily choose the $\lambda_a$ and consider a trajectory
evolved from some $\hat{\omega}$ then another trajectory evolved by a
different choice of $\lambda_a$ from the same initial point, would be
obtained by making a gauge transformation. Clearly, if we identified
points in $\Sigma$ which are related by gauge transformations as being
`physically the same', then we regain determinism in the sense that {\em
physical states} evolve into {\em unique} physical states. Thus {\em the
apparent dynamical indeterminism implied by the first class constraints
appearing in the Hamiltonian, can be resolved by defining equivalence
classes of points of $\Sigma$ under transformations generated by the
constraints (i.e.  gauge transformations) to represent the physical
states of the system.} Notice that this identification of physical
states with equivalence classes under gauge transformations involves
only those first class constraints which appear in the Hamiltonian since
only these have a bearing on the dynamical evolution. 

Since there are $k$ first class constraints, the set of points of
$\Sigma$ which are related by gauge transformations is parameterized by
$k$ parameters and hence the {\em set of gauge equivalence classes} is
parameterized by $2n - k - k$ parameters. This space is called the {\em
Reduced Phase Space}.  This space be made explicit by introducing
additional $k$ `constraints' (now called as {\em gauge fixing
conditions} -- $\chi_a(\omega) \approx 0$). These are required to be
such that the matrix of Poisson brackets of the $\phi^a, \chi_b$
constraints is non-singular. Demanding their preservation in time fixes
the Lagrange multipliers and hence the name.

Having clarified the identification of physical states, the definition
of physical observables and notions of symmetry obviously must be
formulated for physical states. The observables must have unique
evolutions since by definition these are supposed to be functions of
physical states. Only first class functions satisfy this property and
only these qualify to be termed as physical observables. Likewise,
notion of symmetry should refer to invariance with respect to
transformations of {\em physical states}, the generators of
infinitesimal symmetries must map the entire gauge equivalence classes
among themselves and of course preserve the evolution. Clearly these
again have to be first class functions and must additionally Poisson
commute with the Hamiltonian.

To summarize:
\begin{enumerate}
\item Hamiltonian systems with at least one first class constraint, {\em
require} identification of physical states {\em not with individual
points} of the constraint surface {\em but with gauge equivalence
classes of points} of the constraint surface. 

\item In view of the above, it is common to refer to the original phase
space as the {\em kinematical phase space}, $\Gamma_{\mathrm{kin}}$. The
constraint sub-manifold of $\Gamma_{\mathrm{kin}}$ is $\Sigma$. The
physical state space (or reduced phase space) is denoted as
$\Gamma_{\mathrm{phys}} := \Sigma/\sim$ where $\sim$ refers to the gauge
transformations. Note that the physical state space is {\em not} a
sub-manifold of $\Sigma$.

\item Although both the constraints and `conserved quantities' serve to
confine the trajectories the two are distinguished by the fact that
constraint impose limitation on possible initial conditions (restriction
to $\Sigma$) as well as force a non-trivial identification of physical
states to ensure determinism of dynamics. Notions of conserved
quantities, symmetries become meaningful {\em only after} this
identification. Also conserved quantities do not impose any {\em ab
initio} limitation on the possible initial conditions but only on a
subsequent trajectory.

\end{enumerate}

Observe that in the light of the discussion of symmetry and gauge
transformations, a first class Hamiltonian $H_0$ generates a {\em
symmetry transformation}, namely time translations. However there are
theories in which $H_0 = 0$ and $H$ is made up entirely of first class
constraints.  Now the `time evolution' itself becomes a gauge
transformation and hence `no evolution' of physical states. Next chapter
discusses this case.

\chapter{Systems with the Hamiltonian as a constraint}

Consider special types of gauge theories in which $H_0 = 0$ i.e. the
Hamiltonian is entirely made up of first class constraints. The prime
physical example of such a system is the dynamics of {inhomogeneous
cosmological space-times} within the context of Einstein's theory of
General Relativity \footnote{For space-times corresponding to compact
objects, typically asymptotically flat space times, there is a `true
Hamiltonian', the analogue of $H_0$, generating asymptotic time
translations.}. 

As is well known, Einstein's general relativistic theory of gravity (GR)
has a four dimensional manifold on which is defined some metric tensor
(of Minkowskian signature) field which makes it in to a space-time. The
metric tensor is not a fixed entity, as in the case of special
relativity (the Minkowski space-time), but is a dynamical entity i.e. is
determined in conjunction with the (interacting) matter distribution on
the manifold. The equation determining the metric is the Einstein
equation, which is a set of 10, local, partial differential equations of
order 2, for the 10 components of the metric tensor, $g_{\mu\nu}$. It is
non-trivial fact that these equations admit a well-posed initial value
formulation i.e. (i) one can take the 4 dimensional manifold as
$\mathbb{R} \times \Sigma_3$, (ii) specify two symmetric tensor fields,
$g_{ij}, $ and its time derivative, $\dot{g}_{ij}$ on $\Sigma_3$, then
the space-time can be determined for other times provided the `initial
data' satisfies certain conditions. Furthermore, the system of equations
can be cast in the form of constrained Hamiltonian system, with
Hamiltonian given entirely by first class constraints. There are 4 sets
of constraints (per point, since one is dealing with a field theory),
three of which, called {\em vector or diffeomorphism constraints} and
the remaining one called the {\em scalar or Hamiltonian constraint}. The
vector constraints generate usual diffeomorphisms of $\Sigma_3$ while the
scalar constraint generates evolution of the `spatial manifold',
$\Sigma_3$ in the 4-manifold constructing a solution (space-time) of the
Einstein equation. The upshot is that {\em solution space-times of
Einstein equation can be viewed as phase space Hamiltonian trajectories
in the gravitational phase space with initial data satisfying a set of
first class constraints} and with the Hamiltonian given as linear
combination of these constraints. The constraints in the Hamiltonian
formulation of general relativity, reflect the 4-diffeomorphism
invariance of GR \cite{Wald}.

This is pretty complicated to deal with in general, however a
simplification is possible. If we restrict ourselves do the dynamics of
only a special class of 3-geometries, namely, those metric tensors whose
dependence on spatial coordinates (coordinates on $\Sigma_3$) is
completely fixed in a particular manner and only time dependence is to
determined, then GR simplifies to a Hamiltonian system for a finitely
many degrees of freedom (6 for general homogeneous geometries, 3 for
so-called diagonalized models and 1 for homogeneous, isotropic geometry
of the Friedmann-Robertson-Walker (FRW) cosmology) with only the {\em
single} Hamiltonian constraint remaining. We will not need to take any
specific model to illustrate the issues.

We can also {\em construct} systems in which Hamiltonian is the single
(and hence first class) constraint. To see this, let us begin with a
usual {\em un-constrained} Hamiltonian system with a phase space
$\hat{\Gamma}$ and a Hamiltonian $H_0(\omega)$. Let us extend this phase
space to $\Gamma$ by adding two more conjugate variables, $\tau, \pi$.
Let $\phi(\tau, \pi, \omega) := H_0(\omega) - \pi$ be chosen as the
Hamiltonian on $\Gamma$ and take it as a constraint as well, i.e. $H :=
\lambda \phi \approx 0$. Clearly, any evolution (with respect to $T$)
generated by the Hamiltonian is a gauge transformation and therefore
unphysical. Functions which are insensitive to the evolution are the first
class ones (Dirac observables), $f$. Let us look for Dirac observables.
\begin{equation}\label{Dirac}
\{f, H\} ~\approx~ 0 ~~\Rightarrow~~ \{f(\tau, \pi, \omega), H_0(\omega)\}
- \frac{\partial f}{\partial \tau} ~\approx~ 0 \ .
\end{equation}
where we have used the definition of Poisson bracket in the second
equation.

It is clear that functions which are independent of $\tau, \pi$ and
satisfying the usual un-constrained evolution equation in
$\hat{\Gamma}$, {\em are} Dirac observables of the extended constrained
dynamics. Functions depending {\em only} on $\tau$ are not Dirac
observables while those depending only on $\pi$ are Dirac observables.
Constants with respect to the $\tau$-dynamics, also are Dirac
observables. 

With this construction, we see that usual unconstrained dynamics can be
viewed (albeit trivially) as a constrained dynamics in a bigger phase
space. Furthermore, all functions on the unconstrained phase space,
evolving by the un-constrained dynamics {\em are} Dirac observables of
the constrained dynamics. The Dirac observables however are {\em
constants} with respect to the $T$-evolution.

Consider now the trajectories of the constrained dynamics. One finds,
\begin{equation}
\frac{d}{dT} \tau ~ = ~ - \lambda ~~,~~
\frac{d}{dT} \pi ~ = ~ 0 ~~,~~
\frac{d}{dT} \omega ~ = ~ \lambda \{\omega, H_0(\omega)\} ~=~
\lambda \frac{\partial}{\partial\tau} \omega\ .
\end{equation}
The last equality is deduced from (\ref{Dirac}) with $f = \omega$. We
can define a new `time', $T'$ by the equation $dT' = \lambda dT$, so
that
\begin{equation}
\frac{d}{dT'} \tau ~ = ~ - 1 ~~,~~
\frac{d}{dT'} \pi ~ = ~ 0 ~~,~~
\frac{d}{dT'} \omega ~ = \frac{\partial}{\partial\tau} \omega\ .
\end{equation}
Note that $T$ evolution is generated by $\lambda \phi$ while $T'$
evolution is generated by $\phi$. Starting with some initial values at
$T' = 0$, one generates the $T'$-trajectories. All the points along
these are related by gauge transformations generated by $\phi$. Thus the
equivalence classes are in one-to-one and on-to correspondence with the
points $\tau = 0, \pi = \hat{\pi}, \omega = \hat{\omega}$. The gauge
orbits which lie on the constraint surface satisfy $\hat{\pi} =
H(\hat{\omega})$ and hence these orbits are completely determined by
points in $\hat{\Gamma}$. Thus the reduced phase space in this case is
just the un-constrained phase space $\hat{\Gamma}$.

This simple construction brings out a few points. There are {\em two}
notions of `evolution' (i) the $T$ or ($T'$) evolution, which is some
times called an {\em external time evolution} and (ii) the $\tau$
evolution which is correspondingly called an {\em internal time
evolution}. From the point of view of the constrained system, $\tau$ is
just one of the degrees of freedom which is singled out because the
constraint had a particularly simple additive form. The $\tau$ evolution
can thus be thought of as evolution of a set of degrees of freedom with
respect to a singled out degree of freedom. The internal time evolution
is thus also called a {\em relational evolution} while the singled out
degree of freedom is called a {\em clock degree of freedom}. The
arbitrary function, $\lambda$ is also called a {\em lapse function} and
its arbitrariness corresponds to the freedom of re-parameterizing the
external time. The Hamiltonian systems resulting from homogeneous
cosmologies of GR, typically get presented in the form of the
constrained system (the constraint however has a different form in
general). The terminology used above is inherited from the GR context.
One can in fact do a more general and systematic analysis of the notions
of external and internal dynamics which my student Golam Hossain and I
have carried out for finite dimensional systems.

While classically, constrained Hamiltonian systems are interesting in
their own right, they become more challenging at the quantum level. As all
the fundamental interactions of standard model and its extensions are
gauge (field) theories, one has to face these challenges. In the
perturbative analysis, one needs to `fix a gauge', in order that
propagators for gauge fields can be defined and then has to show that
the final observable scattering cross-sections are indeed gauge
invariant. When a quantum theory of gravity is attempted, the
understanding of semiclassical approximation becomes quite complicated
especially in a non-perturbative approach.

\chapter{Exercises}

\begin{enumerate}

\item Check that the coordinate transformations generated by Hamiltonian
vector fields leave the symplectic form {\em invariant}.

\item Check that (\ref{Closed}) property is needed to prove the Jacobi
identity for Poisson brackets.

\item For the Maxwell theory, check that the secondary constraint holds
for all times {\em without} having to require any further constraint.
Furthermore, the Poisson bracket of the two constraints is also zero.

\item Show that the Hamilton's equations of motion can be identified
with the Maxwell equations. 

\item For the Maxwell theory, we have already shown that the canonical
transformations generated by the first class constraints are indeed the
usual gauge transformations (hence in fact the name). Show, by direct
computation, that $F_{\mu\nu}$ are first class functions.

\item Consider a massive, relativistic particle with action $S = m_0\int
d\tau \sqrt{\eta_{\mu\nu} \dot{x}^{\mu} \dot{x}^{\nu}}$. Carry out the
constraint analysis and give examples of first class functions.

\item Consider the four dimensional phase space with coordinates $(q^1,
q^2, p_1, p_2)$. Consider two constraints $\phi(q, p) := p_1^2 + p_2^2 +
(q^1)^2 + (q^2)^2 - R^2$ and $\chi(q, p) := p_2$. Let $H_0(q, p)$ be
some suitable Hamiltonian (not given explicitly) such that these two
constraints are preserved \cite{Radhika}. 

\begin{enumerate}

\item Identify the constraint surface.

\item Compute the expression for the Dirac bracket.

\end{enumerate}

\end{enumerate}


\begin{thebibliography}{99}
%
\bibitem{Treatise} J. G. Papastavridis, {\em Analytical Mechanics: A
Comprehensive Treatise on Dynamics of Constrained Systems}, Oxford, 2002

\bibitem{Dirac} P. A. M. Dirac, {\em Lectures on Quantum Mechanics},
Yeshiva University Press, New York, 1964;

\bibitem{Sudarshan} G. Sudarshan and N. Mukunda, {\em Classical
Dynamics: A Modern Perspective}, chap. 8, 9,

\bibitem{Wald} R. M. Wald, {\em General Relativity}, The University of
Chicago Press, 1984, especially, chapter 10 and appendix E.

\bibitem{Radhika} Radhika Vathsan, {\em J. Math. Phys.} {\bf 37} (1996)
1713-1723; Erratum-ibid. 37 (1996) 6590; hep-th/9507066.
%
\end{thebibliography}
\end{document}